\pdfoutput=1
\documentclass[aps,prd,amsmath,floats,floatfix, twocolumn,
superscriptaddress,nofootinbib,showpacs,longbibliography]{revtex4-1}
\usepackage[T1]{fontenc}
\usepackage[utf8]{inputenc}
\usepackage{lmodern}
\usepackage{verbatim}
\usepackage[dvipsnames, usenames]{xcolor}
\definecolor{linkcolor}{rgb}{0.0,0.3,0.5}
\usepackage[hypertexnames=false, unicode, colorlinks=true, linkcolor=linkcolor,
citecolor=linkcolor, filecolor=linkcolor,urlcolor=linkcolor,
pdfusetitle]{hyperref}
\usepackage[all]{hypcap}
\usepackage{graphicx}
\usepackage{xspace}
\usepackage{amssymb}
\usepackage{microtype}

\usepackage{url}

\usepackage[english]{babel}
\usepackage{blindtext}

\usepackage[normalem]{ulem} 
\usepackage{bm} 
\usepackage{mathrsfs}

\DeclareMathAlphabet{\mathpzc}{OT1}{pzc}{m}{it}

\usepackage[nomessages]{fp}

\begin{document}
\title{Mapping between black-hole perturbation theory and numerical relativity II: gravitational-wave momentum}
\newcommand{\KITP}{\affiliation{Kavli Institute for Theoretical Physics,\\University of California Santa Barbara, Kohn Hall, Lagoon Rd, Santa Barbara, CA 93106}}

\author{Tousif Islam}
\email{tislam@kitp.ucsb.edu}
\KITP

\hypersetup{pdfauthor={Islam et al.}}

\date{\today}

\begin{abstract}
We report an approximate, non-trivial mapping of angular (linear) momentum in gravitational waves obtained from numerical relativity (NR) and adiabatic point-particle black hole perturbation theory (BHPT) in the comparable mass regime for quasi-circular, non-spinning binary black holes. 
This mapping involves two time-independent scaling parameters, $\alpha_{J}$ ($\alpha_{P}$) and $\beta_{J}$ ($\beta_{P}$), that adjust the BHPT angular (linear) momentum and the BHPT time respectively such a way that it aligns with NR angular (linear) momentum. 
Our findings indicate that this scaling mechanism works really well until close to the merger.
In addition to the comparison of $\alpha_{J}$ ($\alpha_{P}$) with relevant values obtained from the waveform and flux scalings, we explore the mass ratio dependence of the scaling parameter $\alpha_{J}$ ($\alpha_{P}$). Finally, we investigate their possible connection to the missing finite size correction for the secondary black hole within the BHPT framework and the implication of these scalings on the remnant properties of the binary.
\end{abstract}

\maketitle

\section{Introduction}
Gravitational radiation from the merger of a binary black hole (BBH) is usually written as a superposition of $-2$ spin-weighted spherical harmonic modes with indices $(\ell,m$):
\begin{align}
h(t,\theta,\phi;\boldsymbol{\lambda}) &= \sum_{\ell=2}^\infty \sum_{m=-\ell}^{\ell} h^{\ell m}(t;\boldsymbol\lambda) \; _{-2}Y_{\ell m}(\theta,\phi)\,,
\label{hmodes}
\end{align}
where $\boldsymbol{\lambda}$ is the set of intrinsic parameters (such as the masses and spins of the binary) describing the binary, $\theta$ is the polar angle, and $\phi$ is the azimuthal angle.

Simulations of BBH mergers help us understand the dynamics of black holes and the phenomenology of waveforms. Two of the primary numerical frameworks employed in simulating BBH mergers are numerical relativity (NR)~\cite{Mroue:2013xna,Boyle:2019kee,Healy:2017psd,Healy:2019jyf,Healy:2020vre,Healy:2022wdn,Jani:2016wkt,Hamilton:2023qkv} and adiabatic point-particle perturbation theory (BHPT)~\cite{Sundararajan:2007jg,Sundararajan:2008zm,Sundararajan:2010sr,Zenginoglu:2011zz,Fujita:2004rb,Fujita:2005kng,Mano:1996vt,throwe2010high,OSullivan:2014ywd,Drasco:2005kz}.
NR simulations solve the nonlinear Einstein equations without resorting to any approximations. This approach proves to be effective for comparable mass binaries, typically within the range $1 \leq q \leq 10$, where $q:=m_1/m_2$ denotes the mass ratio of the binary. Here, $m_1$ and $m_2$ are the masses of the larger and smaller black holes, respectively. As the binary enters the intermediate mass ratio regime ($10 \leq q \leq 100$), due to the multi-scale physics, algorithmic complexity increases significantly making these simulations extremely challenging.

On the other hand, the BHPT framework linearizes the Einstein equation, treating smaller black hole to be a point particle~\cite{Sundararajan:2007jg, Sundararajan:2008zm, Sundararajan:2010sr, Zenginoglu:2011zz, Fujita:2004rb, Fujita:2005kng, Mano:1996vt, throwe2010high, OSullivan:2014ywd, Drasco:2005kz}. As a result, this framework provides accurate simulations only in the extreme mass ratio limit (i.e. for $q \to \infty$ or, in most practical purposes, when $q \ge 10^4$). As the binary moves towards the intermediate mass ratio and comparable mass regime, these simulations become less reliable. A number of recent advances however have extended the regime of validity of both NR and BHPT further~\cite{Pound:2021qin,Miller:2020bft,Wardell:2021fyy,Islam:2022laz,Rifat:2019ltp,Islam:2022laz,Islam:2023mob,Islam:2023qyt,Islam:2023aec,Islam:2023jak,Lousto:2020tnb, Lousto:2022hoq}. 

Over the years, significant efforts have been devoted to understanding the similarities and differences between BHPT and NR simulations in the comparable mass and intermediate mass ratio regime~\cite{Lousto:2010tb, Lousto:2010qx, Nakano:2011pb, NavarroAlbalat:2022tvh, Albalat:2022lfz, Ramos-Buades:2022lgf, vandeMeent:2020xgc, LeTiec:2014oez, LeTiec:2013uey, LeTiec:2011dp, LeTiec:2011ru}. These studies not only enhance our interpretation of binary dynamics in greater detail but also illuminate the higher-order correction terms required in the BHPT framework~\cite{NavarroAlbalat:2022tvh, Albalat:2022lfz, Ramos-Buades:2022lgf, vandeMeent:2020xgc}.

One notable advancement is the reporting of an approximate mapping, named the $\alpha$-$\beta$ scaling, between linear BHPT and nonlinear NR waveforms in the comparable mass regime. For any given mass ratio, the mapping reads~\cite{Islam:2022laz}:
\begin{align} \label{eq:EMRI_rescale}
h^{\ell,m}_{\tt NR}(t_{\tt NR} ; q) \sim {\alpha_{\ell}} h^{\ell,m}_{\tt BHPT}\left( \beta t_{\tt BHPT};q \right) \,,
\end{align}
where, $h^{\ell,m}_{\tt NR}$ and $h^{\ell,m}_{\tt BHPT}$ represent the NR and BHPT waveforms, respectively, as functions of the NR time $t_{\tt NR}$ and BHPT time $t_{\tt BHPT}$ respectively. The scaling parameters $\alpha$ depends on the $\ell$ value of the mode while $\beta$ is same for all modes.
Once the mapping is applied, the scaled BHPT waveforms exhibit excellent agreement with NR in the comparable mass regime, exhibiting errors of approximately $10^{-3}$ or less in the quadrupolar mode~\cite{Islam:2022laz}. Furthermore, the scaled BHPT waveforms match with a set of recent high mass ratio NR data ranging from $q=15$ to $q=128$~\cite{Islam:2023qyt}. Although the exact origin of this mapping is not fully known, further studies provide convincing evidence that the scaling parameters $\alpha_\ell$ and $\beta$ mostly account for the missing finite size of the smaller black hole in the BHPT framework~\cite{Islam:2023aec}, possibly along with other effects such as post-adiabatic corrections.

It has been further noted that a similar mapping also exists between NR and BHPT fluxes, taking the following form for a given mass ratio~\cite{Islam:2023csx}:
\begin{equation}
    \mathcal{F}_{\tt NR} (t_{\tt NR}) =  \alpha_{\mathcal{F}} \times \mathcal{F}_{\tt BHPT} (\beta_{\mathcal{F}} \times t_{\tt BHPT}),
    \label{eq:flux_scaling}
\end{equation}
where $\mathcal{F}_{\tt NR}$ and $\mathcal{F}_{\tt BHPT}$ are the gravitational-wave fluxes calculated using NR and BHPT respectively while $\alpha_{\mathcal{F}}$ and $\beta_{\mathcal{F}}$ are the scaling parameters.

Here, we investigate whether such a mapping also extends to other derived quantities, such as the linear and angular momentum of the binary, in the comparable mass regime. Understanding these potential scalings will greatly aid future binary modeling with BHPT frameworks. We obtain the NR data from the NR catalog maintained by the SXS collaboration (\url{https://data.black-holes.org/waveforms/catalog.html})~\cite{Mroue:2013xna,Boyle:2019kee}. In particular, we utilize the following  13 NR simulations: $q=3$ (\texttt{SXS:BBH:2265}), $q=3.5$ (\texttt{SXS:BBH:0193}), $q=4$ (\texttt{SXS:BBH:1220}),  $q=4.5$ (\texttt{SXS:BBH:0295}), $q=5.0$ (\texttt{SXS:BBH:0107}), $q=5.5$ (\texttt{SXS:BBH:0296}), $q=6$ (\texttt{SXS:BBH:0181}), $q=6.5$ (\texttt{SXS:BBH:0297}), $q=7$ (\texttt{SXS:BBH:0298}), $q=7.5$ (\texttt{SXS:BBH:0299}), $q=8.0$ (\texttt{SXS:BBH:0063}), $q=9.2$ (\texttt{SXS:BBH:1108}) and $q=10.0$ (\texttt{SXS:BBH:1107}). 
On the other hand, we generate BHPT data for these mass ratios using the \texttt{BHPTNRSur1dq1e4} ~\cite{Islam:2022laz} model from the \texttt{BHPTNRSurrogate(s)~\cite{BHPTSurrogate}}) package available in the \texttt{Black Hole Perturbation Toolkit}~\cite{BHPToolkit}. This is a reduced-order surrogate model trained on waveform data generated with a time-domain Teukolsky equation solver~\cite{Sundararajan:2007jg,Sundararajan:2008zm,Sundararajan:2010sr,Zenginoglu:2011zz}. 

The rest of the paper is structured as follows. In Section~\ref{sec:angmom_mapping}, we delve into the mapping between NR and BHPT angular momentum. Section~\ref{sec:mapping_linmom} then explores a similar mapping between NR and BHPT linear momentum. We proceed to compare various scalings obtained in quantities such as waveform modes, fluxes, angular momentum, and linear momentum in Section~\ref{sec:understanding_mapping}. We discuss the potential implication of our results on the remnant properties of the binary in Section~\ref{sec:remnant}. Finally, in Section~\ref{sec:conclusion}, we outline future directions.

\section{Mapping between BHPT and NR  angular momentum}
\label{sec:angmom_mapping}
In this section, we explore the connection between BHPT and NR angular momentum. We first present the framework used in computing the angular momentum evolution. We then report the observed mapping in Section~\ref{sec:alpha_beta_angmom}. We demonstrate the mapping at $q=4$ in Section~\ref{sec:angmom_q4}. Subsequently, we discuss the mass ratio dependence of the observed mapping in Section~\ref{sec:mass_ratio_dependence_angmom}.

We calculate the angular momentum directly from the waveforms $h^{\ell m}(t)$. These waveforms are aligned such that $t=0$ indicate merger~\footnote{We denote the time corresponding to the maximum amplitude of the $(2,2)$ mode as the time of merger.}. We point out that the initial mass-scale of the NR and BHPT simulations are different. In particular, NR uses the total mass of the binary $M$ while BHPT simulations use the mass of the primary, $m_1$. Before computing the angular momentum, we therefore scale the BHPT waveform and time with the mass-scale transformation factor of $\frac{1}{1+1/q}$ to ensure both have the same mass-scale.

The rate of loss of angular momentum during binary evolution is given by~\cite{Gerosa:2018qay}:
\begin{align}
\frac{d J_x}{dt} =  &\lim_{r\rightarrow\infty} \frac{r^2}{32 \pi} \:
\Im \Bigg[ \sum_{\ell m} \,h^{\ell m}
\Big( f_{\ell m}\, \dot{\bar{h}}^{\ell,m+1}
\notag \\&+ f_{\ell,-m}\, \dot{\bar{h}}^{\ell,m-1} \Big) \Bigg]\;  ,
\label{eq:dt_jx}\\
\frac{d J_y}{dt} = &- \lim_{r\rightarrow\infty} \frac{r^2}{32 \pi} \:
\Re \Bigg[ \sum_{\ell,m} \, h^{\ell m}  \Big( f_{\ell m}\, \dot{\bar{h}}^{\ell,m+1}
\notag \\ &- f_{\ell,-m}\, \dot{\bar{h}}^{\ell,m-1} \Big) \Bigg]\; ,
\label{eq:dt_jy} \\
\frac{d J_z}{dt} =  &\lim_{r\rightarrow\infty} \frac{r^2}{16 \pi} \:
\Im \Bigg[ \sum_{\ell,m} \,m\, h^{\ell m}
 \,\dot{\bar{h}}^{\ell m} \Bigg] \; ,
\label{eq:dt_jz}
\end{align}
where
\begin{eqnarray}
f_{\ell m} = \sqrt{\ell(\ell+1) - m(m+1)} \; .
\label{flm}
\end{eqnarray}
Here, $\bar{h}$ is the complex conjugate of $h$ and $\dot{h}$ denotes time derivative of the waveform.
To compute the total loss of the angular momentum since the start of the waveform, we integrate $d\mathbf{J}/dt$ where $\mathbf{J}:=[J_x, J_y, J_z]$. For non-spinning binaries, $J_x$ and $J_y$ are negligible; therefore, we only focus on $J_z$. We obtain:
\begin{equation}
    J_z = \int_{t_{\rm initial}}^{t_{\rm final}} \frac{d J_z}{dt} dt,
\end{equation}
where $t_{\rm initial}$ and $t_{\rm final}$ are the start and end time of the input waveform $h^{\ell m}(t)$. 

Using the above-mentioned framework, we compute the angular momentum $J_z$ for both NR and BHPT from $h^{\ell,m}_{\tt NR}(t_{\tt NR})$ and $h^{\ell,m}_{\tt BHPT}(t_{\tt BHPT})$ respectively using \texttt{gw\_remnant}~\cite{Islam:2023mob,gwremnant}. We denote them as $J_{z, \tt NR}$ and $J_{z, \tt BHPT}$ respectively.

For comparison, we also compute the angular momentum using a post-Newtonian expression that reads (see Eq.~2.36 of Ref.~\cite{LeTiec:2011ab}):
\begin{equation}\label{eq:PN_angmom}
J_{\rm PN} =  \frac{\nu}{x^{1/2}}\left( 1 + j_{1}(\nu) x + j_{2}(\nu) x^2 + j_{3}(\nu) x^3 \right),
\end{equation}
where
\begin{align}
j_{1}(\nu) &= \left( \frac{3}{2} + \frac{\nu}{6}\right) \; ,\\
j_{2}(\nu) &=  \left( \frac{27}{8} - \frac{19\nu}{8} + \frac{\nu^2}{24}\right) \; ,\\
j_{3}(\nu) &=  \left( \frac{135}{16} + \left[ -\frac{6889}{144} + \frac{41\pi^2}{24}\right] \nu + \frac{31\nu^2}{24} + \frac{7\nu^3}{1296} \right) \;.
\label{eq:J0}
\end{align}
Here, $\nu=\frac{q}{(1+q)^2}$ is the symmetric mass ratio and $x=\omega^{2/3}$ is the dimensionless frequency parameter with $\omega$ being the orbital frequency.

\subsection{$\alpha$-$\beta$ scaling of the angular momentum}
\label{sec:alpha_beta_angmom}
We notice that, for a given mass ratio, BHPT and NR angular momentum values exhibit the following $\alpha$-$\beta$ mapping:
\begin{equation}
    {J}_{z,\tt NR} (t_{\tt NR}) =  \alpha_{{J}} \times {J}_{z,\tt BHPT} (\beta_{{J}} \times t_{\tt BHPT}),
    \label{eq:angmom_scaling}
\end{equation}
where $\alpha_{{J}}$ and $\beta_{{J}}$ are the time-independent scaling parameters.

\begin{figure}
\includegraphics[width=\columnwidth]{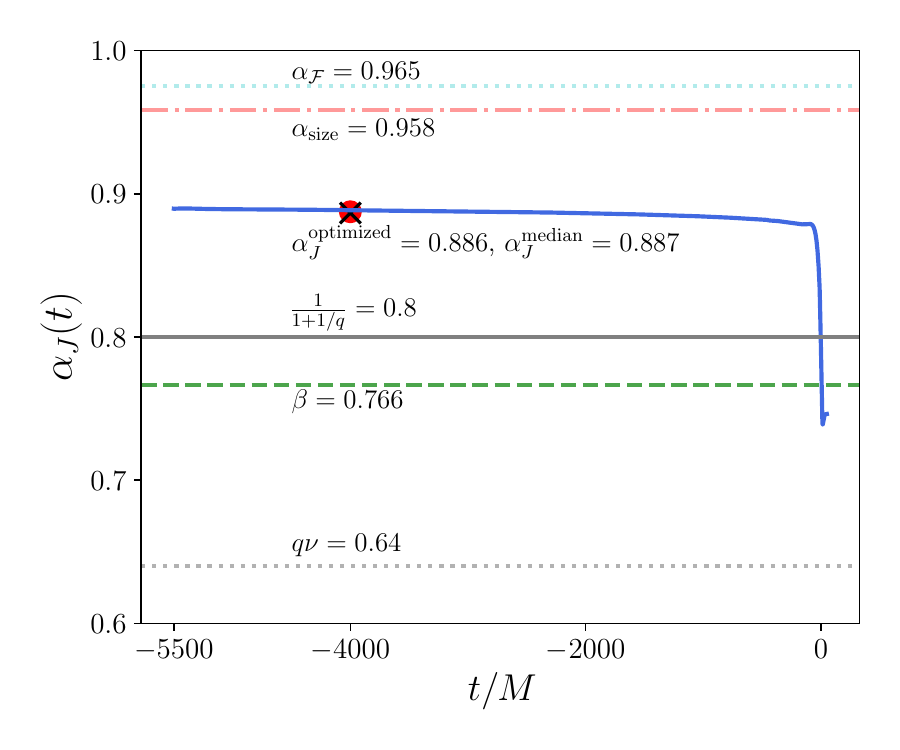}
\caption{We show the temporal variation of the $\alpha_{J}(t)$ (blue solid line) for a mass ratio of $q=4$. Additionally, we show the naive mass-scale transformation factor $\frac{1}{1+1/q}$ (grey solid line), $\beta$ obtained in Ref.~\cite{Islam:2022laz} (green dashed line) and $\alpha_{\rm size}$ obtained in Ref.~\cite{Islam:2023qyt} (dash-dotted red line). We further show $\alpha_{J}^{\rm optimized}$, obtained through loss-function minimization, as a red circle and $\alpha_{J}^{\rm median}$ as a black cross. More details are in Section \ref{sec:angmom_q4}.}
\label{fig:q4_alpha_timeseries_angmom}
\end{figure}

We must note that the values of $\alpha_{J}$ and $\beta_{J}$ are not known beforehand. However, it is possible to make a reasonable guess for $\beta_{J}$. It is reasonable to assume that the time scaling will be the same for the waveform, flux, and the angular momentum. In other words, the $\alpha$-$\beta$ mapping for the angular momentum should inherit the same value of $\beta$ that works for the waveform and the flux. 
However, there is a small catch. In the waveform scaling, the mass scale for NR was $M$, whereas the mass scale for BHPT data was $m_1$~\cite{Islam:2022laz}. However, in this paper, we have already ensured that both NR and BHPT use the same mass scale of $M$. Consequently, $\beta_{J}$ (and $\beta_{\mathcal{F}}$) will be equal to $\beta$ only when the mass-scale transformation factor $\frac{1}{1+1/q}$ is factored out of the $\beta$ parameter:
\begin{equation}
    \beta_{J} := \beta_{\mathcal{F}} = \beta \times (1+1/q),
\end{equation}
where $\beta$ is computed using the following analytical approximation~\cite{Islam:2022laz}:
\begin{align}
\begin{split}
\beta(q) =  1 & - \frac{1.238}{q} + \frac{1.596}{q^2} 
- \frac{1.776}{q^3} + \frac{1.0577}{q^4}.\;
\end{split}
\label{beta_fit}
\end{align}
We can compute $\alpha_J$ in two different ways. First, we can treat the computation of $\alpha_J$ as an optimization problem and find the optimal value of $\alpha_J$ that minimizes the following loss function:
\begin{equation}
   \mathcal{E}(\alpha_{J}) = \int \frac{{J}_{z,\tt NR} (t_{\tt NR}) - \alpha_{{J}} {J}_{z,\tt BHPT} (\beta_{{J}} t_{\tt BHPT})}{{J}_{z,\tt NR} (t_{\tt NR})}.
\end{equation}
We denote this as $\alpha_{J}^{\rm optimized}$. At this point, it is worth mentioning that while the scaling parameters $\alpha_{{J}}$ and $\beta_{{J}}$ are intended to have no temporal evolution, we can, in principle, obtain their time variation for a given mass ratio by computing the ratio of the NR and BHPT angular momentum:
\begin{equation}
    \alpha_{J}(t_{\tt NR}) = \frac{J_{\tt BHPT} (\beta_{J} t_{\tt BHPT})}{J_{\tt NR} (t_{\tt NR})}.
    \label{eq:alpha_J_t}
\end{equation}
One can then use the median of the $\alpha_{J}(t_{\tt NR})$ time series as their scaling parameter $\alpha_{J}$. We denote this as $\alpha_{J}^{\rm median}$.

\begin{figure*}
\includegraphics[width=\textwidth]{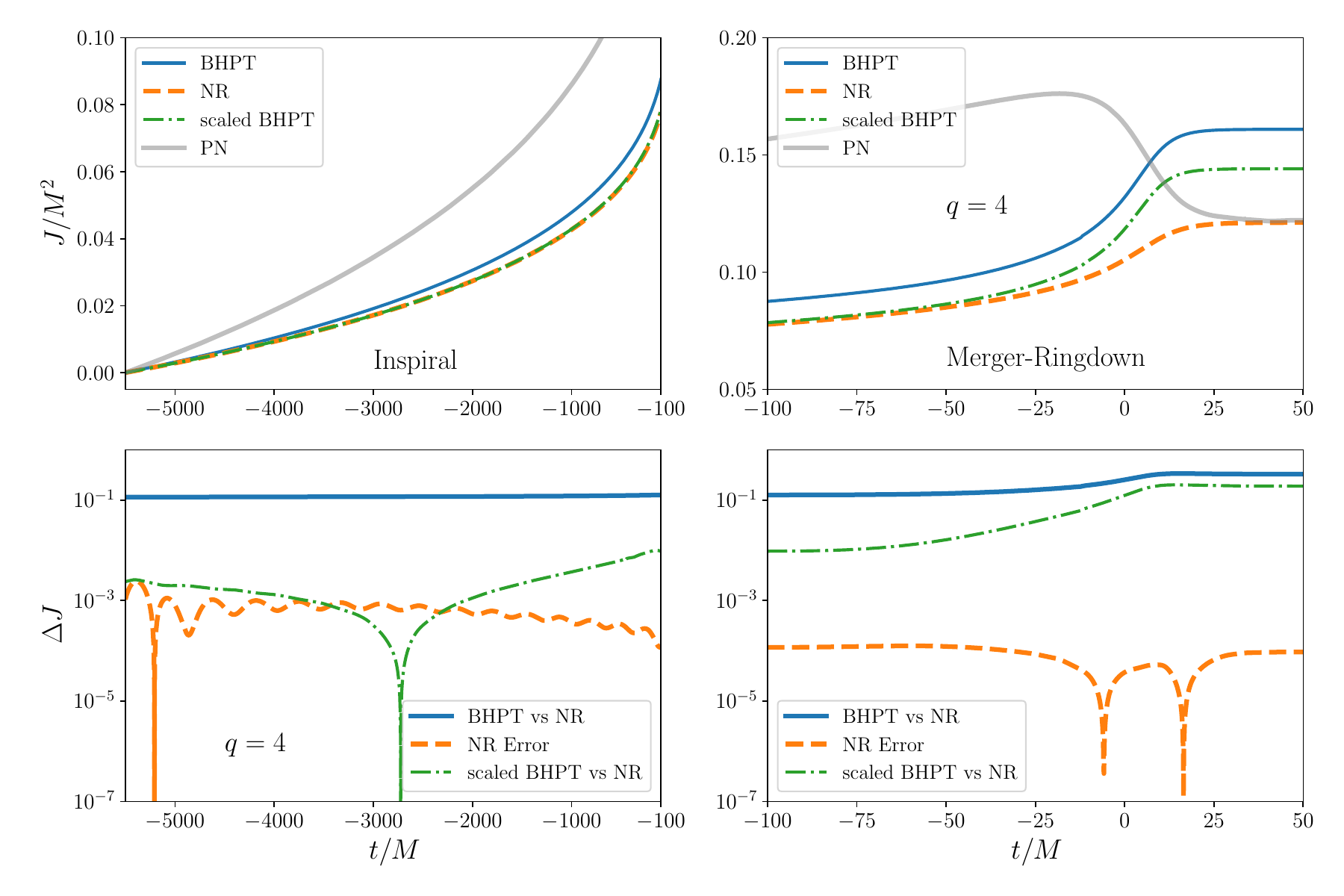}
\caption{\textit{Upper panels: } we show the angular momentum profiles obtained from BHPT (solid blue lines) and NR (dashed orange lines) for a binary with mass ratio $q=4$ along with $\alpha_{J}$-$\beta_{J}$ scaled BHPT angular momentum (using Eq.(\ref{eq:angmom_scaling})) as green dashed-dotted lines. Left panel shows the inspiral part (where the mapping works well) whereas right panel zooms into the merger-ringdown (where the mapping breaks down due to different final mass/spin scale). For comparison, we also show the PN angular momentum as grey solid lines.  For clearer comparison, we set the initial angular momentum to zero.
\textit{Lower panels: } we show the relative error between ((i) BHPT and NR angular momentum (solid blue line) and (ii) $\alpha_{J}$-$\beta_{J}$ scaled BHPT and NR angular momentum (dashed-dotted green line) along with the numerical error in NR simulations (dashed orange line).
More details are in Section \ref{sec:angmom_q4}.}
\label{fig:q4_angmom}
\end{figure*}
\begin{figure}
\includegraphics[width=\columnwidth]{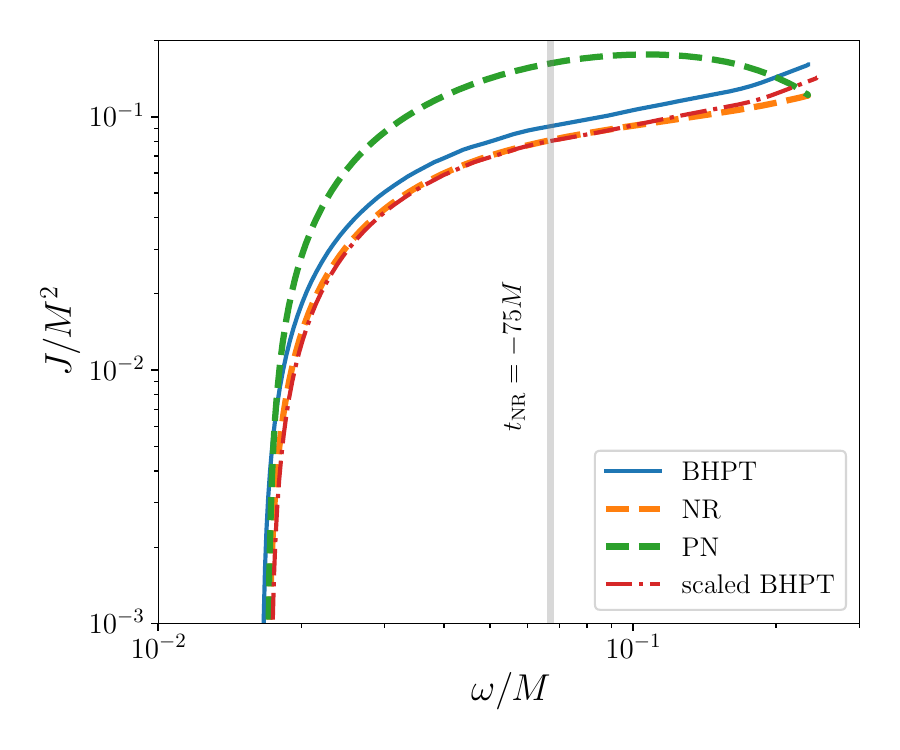}
\caption{We show the angular momentum profiles obtained from NR, BHPT and PN approximation as a function of the orbital frequencies. For comparison, we also show $\alpha_J$-$\beta_J$ scaled BHPT angular momentum. Vertical dashed line indicate $t=-75M$. More details are in Section \ref{sec:angmom_q4}.}
\label{fig:q4_angmom_omega}
\end{figure}

\subsection{Demonstration at $q=4$}
\label{sec:angmom_q4}
We demonstrate the angular momentum mapping at $q=4$, representing a moderate value of mass ratio within the comparable mass regime. This choice ensures that our observations are not biased by any special cases, as may be the case for equal mass binaries. 

Following the methods described in Section~\ref{sec:alpha_beta_angmom}, we computed both the $\alpha_J(t_{\tt NR})$ time series as well as $\alpha_J^{\rm optimized}$. We observe that the values of $\alpha_J^{\rm optimized}$ ($=0.88658$) and $\alpha_J^{\rm median}$ ($=0.88758$), derived from the $\alpha_J(t_{\tt NR})$ time series, are very close to each other. 
In Fig.\ref{fig:q4_alpha_timeseries_angmom}, we present the $\alpha_J(t_{\tt NR})$ time series along with various values of $\alpha$ obtained either through optimization ($\alpha_J^{\rm optimized}$; red circle) or by computing the median of the time series ($\alpha_J^{\rm median}$, black cross). For comparison, we include the mass-scale transformation factor of $\frac{1}{1+1/q} (=0.8)$ and the transformation factor $q\nu$ from mass ratio $q$ to the symmetric mass ratio $\nu=\frac{q}{(1+q)}$. Additionally, we show the $\beta$ parameter used to match the BHPT time (with a mass scale of $m_1$) to NR time (with a mass scale of $M$) from Ref.\cite{Islam:2022laz}, the $\alpha_{\mathcal{F}}$ parameter to scale BHPT fluxes to NR from Ref.\cite{Islam:2023csx}, and $\alpha_{\rm size}$ from Ref.\cite{Islam:2023qyt}, accounting for missing finite size effects/post-adiabatic corrections. We note that the $\alpha_J$ estimates significantly differ for any other parameters listed above.

In Fig.~\ref{fig:q4_angmom}, we show the angular momentum profiles obtained from NR and BHPT. Angular momentum obtained from BHPT is represented by the solid blue line, while the dashed orange line corresponds to NR. For comparison, we also show the $\alpha_{J}$-$\beta_{J}$ scaled BHPT angular momentum as green dashed-dotted lines. 
Additionally, we include the PN angular momentum as grey solid lines (using Eq.(\ref{eq:PN_angmom})). To facilitate a clearer comparison, we project all the angular momentum time series onto the same time grid and set the initial angular momentum to zero. This allows us to disregard the angular momentum evolution until the start of the waveform, enabling a focused examination of the relevant portion of the waveform. 
It is important to note that incorporating the contribution of the angular momentum evolution until the start of the waveform will only introduce a constant offset to the obtained angular momentum profile. This constant offset is often computed with PN and will be the same for both NR, BHPT or PN. 

\begin{figure*}
\includegraphics[width=\textwidth]{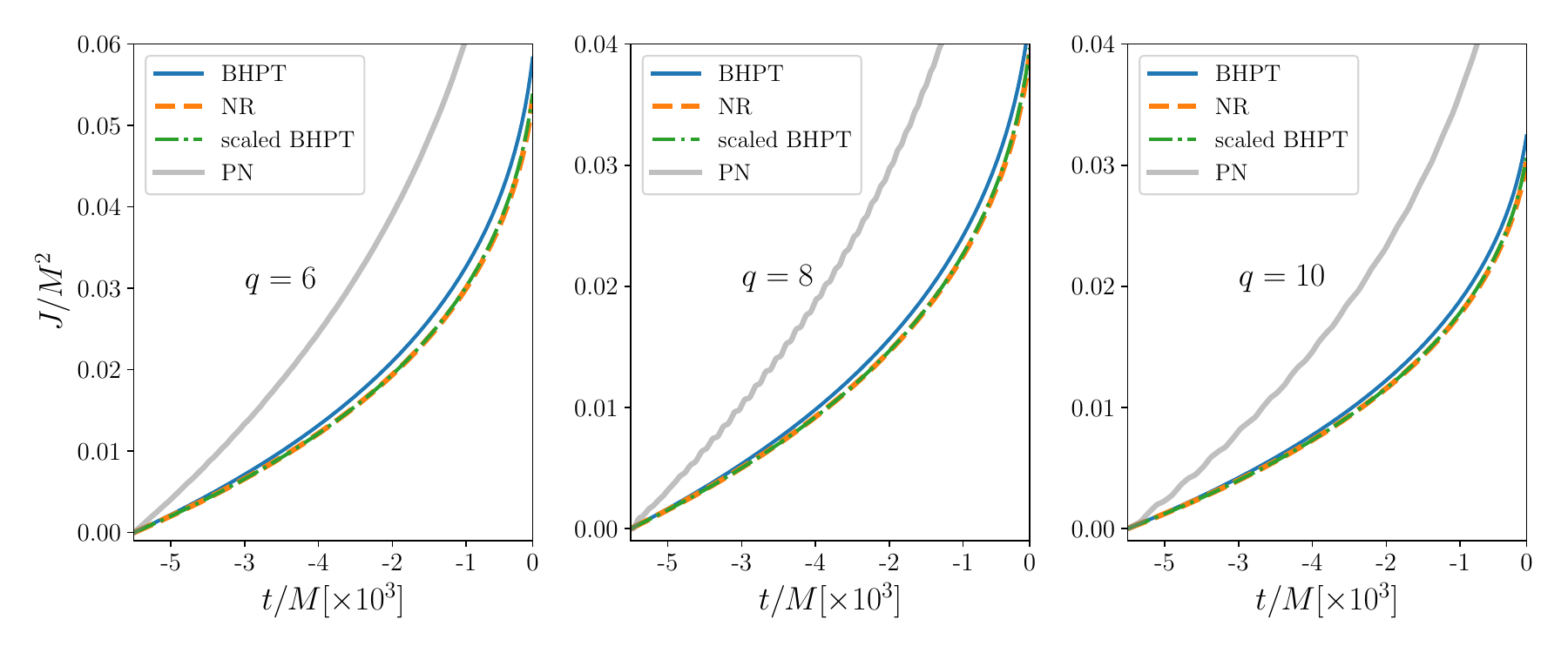}
\caption{We show the angular momentum profiles obtained from BHPT (solid blue lines) and NR (dashed orange lines) along with $\alpha_{J}$-$\beta_{J}$ scaled BHPT angular momentum (using Eq.(\ref{eq:angmom_scaling})) as green dashed-dotted lines for $q=6$ (left panel), $q=8$ (middle panel) and $q=10$ (right panel). For comparison, we also show the PN angular momentum as grey lines. For meaningful comparison, we set the initial angular momentum to zero. We only show the inspiral part, up to $t=-100M$, where the mapping works really well. More details are in Section \ref{sec:mass_ratio_dependence_angmom}.}
\label{fig:q_6_8_10_angmom}
\end{figure*}

\begin{figure}
\includegraphics[width=\columnwidth]{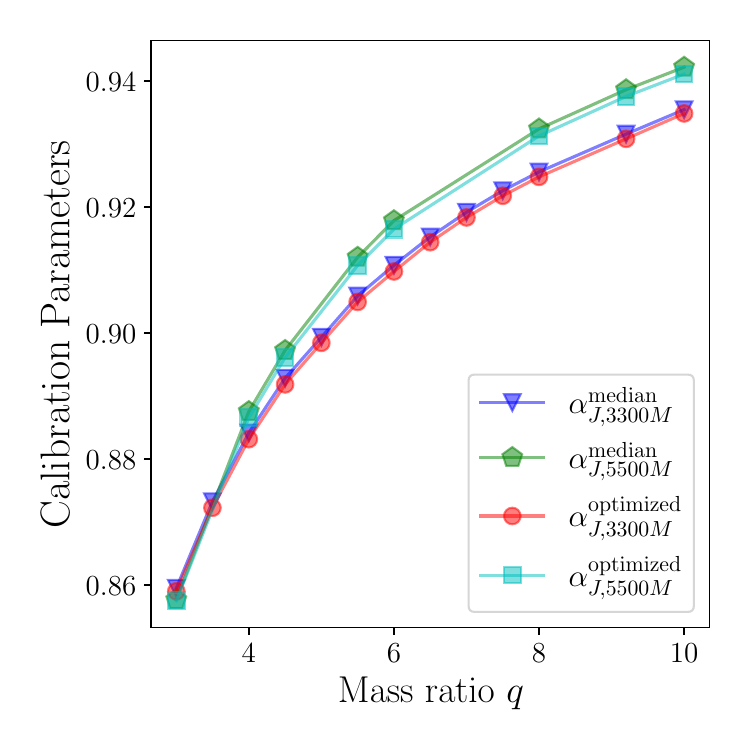}
\caption{We show the calibration parameters $\alpha_{J}$, obtained using different length of the NR data and methods as a function of the mass ratio ranging from $q=3$ to $q=10$. We denote these parameters as $\alpha^{\rm optimized}_{J,5500M}$, $\alpha^{\rm median}_{J,5500M}$, $\alpha^{\rm optimized}_{J,3300M}$ and $\alpha^{\rm median}_{J,3300M}$. Subscript denote whether we use $\sim 3300M$ or $\sim 5500M$ of NR data whereas superscripts denote the method used. More details are in Section \ref{sec:mass_ratio_dependence_angmom}.}
\label{fig:alpha_angular_momentum_vs_q}
\end{figure}

It is evident that the PN values do not match either NR or BHPT in this regime. Notably, the BHPT and NR angular momentum do not align despite sharing the same mass scale. However, adjusting the BHPT angular momentum appropriately through the values of $\alpha_{J}$ and $\beta_{J}$ results in a close match with the NR data. It is also important to highlight that we use the value of $\alpha_J^{\rm optimized}$ as our $\alpha_{J}$. We do observe, however, that the scaling breaks down in the merger-ringdown part, especially after $t=-100M$ (Fig.\ref{fig:q4_angmom}, right panel). This breakdown of the $\alpha$-$\beta$ scaling around the merger was previously observed for both waveform scaling~\cite{Islam:2022laz} and flux scaling~\cite{Islam:2023csx}. The most plausible reason for such a breakdown is the change in mass and spin values for the remnant~\cite{Islam:2023aec}. This happens because while the mass and spin values are known to change during the binary evolution, BHPT simulations fix these values. Therefore, while the initial and remnant mass/spin values are different in NR data, they are typically the same in BHPT simulations.

Next, we assess the effectiveness of the scaling by computing the relative error between NR and the (scaled) BHPT angular momentum (Fig.~\ref{fig:q4_angmom}, lower panels). For comparison, we also show the numerical error in NR simulations. NR errors are calculated by comparing the angular momentum profiles obtained from the highest two resolution NR datasets. We observe that the relative error between the scaled BHPT and NR roughly matches the numerical error in NR itself in the inspiral part up to $t=-2000M$, beyond which it remains around $10^{-2}$ until very close to the merger. This suggests that the scaling is highly effective. However, the relative error between BHPT and NR is approximately $\sim 10^{-1}$ throughout the binary evolution and increases further in the merger-ringdown part.

Lastly, we inspect the angular momentum profiles as a function of the orbital frequencies in Fig.~\ref{fig:q4_angmom_omega}. It shows that while PN approximations quickly deviate from NR, $\alpha_J$-$\beta_J$ scaled BHPT flux can match them reasonably well until close to merger. Furthermore, BHPT fluxes yield better match to NR than PN in most part of the waveform considered here.

\subsection{Mass ratio dependence}
\label{sec:mass_ratio_dependence_angmom}
We repeat our experiment for a total of 13 mass ratio values ranging from $q=3$ to $q=10$. In all cases, we observe that the scaling works remarkably well in the inspiral part and breaks down in the merger-ringdown part. In Fig.~\ref{fig:q_6_8_10_angmom}, we illustrate the BHPT, NR, and $\alpha_J$-$\beta_J$ scaled BHPT angular momentum for three different mass ratio values: $q=[6,8,10]$. As the mass ratio increases, the scale of angular momentum variations over the binary evolution timescale decreases, and so does the differences between NR and BHPT angular momentum. Nonetheless, in all cases, the scaling remains effective.

\begin{figure*}
\includegraphics[width=\textwidth]{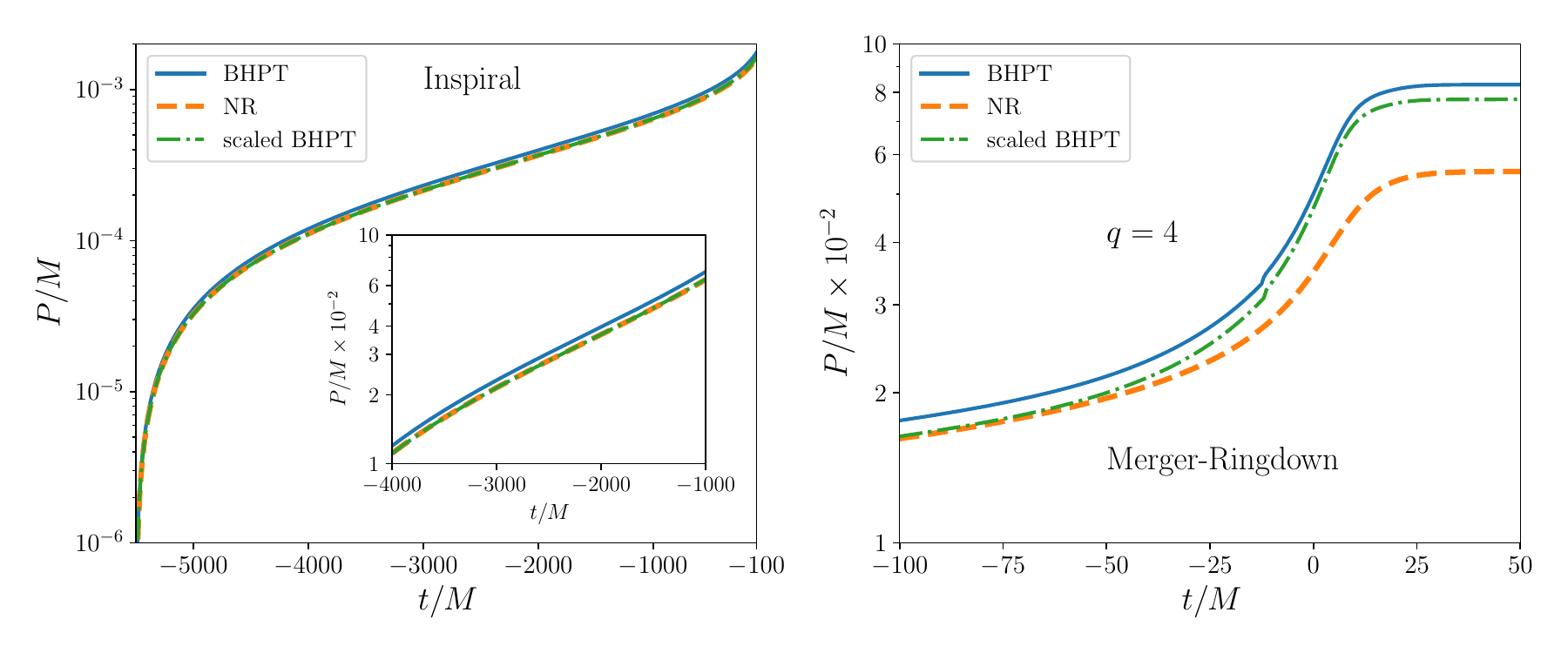}
\caption{We show the linear momentum profiles obtained from BHPT (solid blue lines) and NR (dashed orange lines) for a binary with mass ratio $q=4$ along with $\alpha_{J}$-$\beta_{J}$ scaled BHPT linear momentum (using Eq.(\ref{eq:angmom_scaling})) as green dashed-dotted lines. Left panel shows the inspiral part (where the mapping works extremely well) whereas right panel zooms into the merger-ringdown (where the mapping breaks down due to different final mass/spin scale).
More details are in Section \ref{sec:linmom_q4}.}
\label{fig:q4_linmom}
\end{figure*}

Next, we explore how the value of $\alpha_{J}$ varies with the mass ratio, using a consistent length of NR data across different mass ratios. The common length is determined by the shortest duration NR data. Following the approach in Ref.~\cite{Islam:2023csx}, we conduct this analysis twice. First, we consider all 13 NR datasets with a common duration of $\sim 3300M$ and denote the resulting $\alpha_{J}$ as $\alpha^{\rm median}_{J,3300M}$. Second, we use only 8 NR datasets with a relatively longer duration (approximately $\sim 5500M$) and denote the corresponding value of $\alpha_{J}$ as $\alpha^{\rm median}_{J,5500M}$. For both cases, we also compute the optimized values obtained from the loss minimization process described in Section~\ref{sec:alpha_beta_angmom} and denote them as $\alpha^{\rm optimized}_{J,3300M}$ and $\alpha^{\rm optimized}_{J,5500M}$. Figure~\ref{fig:alpha_angular_momentum_vs_q} shows how these parameters change with the mass ratio. We observe that $\alpha^{\rm median}_{J,5500M}$ tends to take slightly larger values compared to $\alpha^{\rm median}_{J,3300M}$, indicating a subtle distinction between the two based on data length. Similarly, the median values seems to be slightly larger than the corresponding opitmized value of $\alpha_J$.

Finally, we obtain analytical expression of $\alpha_{J}$ (for which we use $\alpha^{\rm optimized}_{J,3300M}$) by performing fits in terms of $q$ (using the \texttt{scipy.optimize.curve\_fit}~\cite{scipyfit} module) and obtain:
\begin{align}
\alpha^{\rm median}_{J}(q) \approx & 1 -0.11849193 \times (\frac{1}{q}) +0.03480474 \times (\frac{1}{q})^2\notag\\
& -0.00375373 \times (\frac{1}{q})^3 +0.0001398 \times (\frac{1}{q})^4.
\end{align}
We further obtain the fit as a function of the symmetric mass ratio $\nu$ as:
\begin{align}
\alpha^{\rm median}_{J}(\nu) \approx & 1 -0.62314124 \times \nu -2.61924118 \times \nu^2\notag\\
& +22.81600927  \times \nu^3 -66.54634884 \times \nu^4.
\end{align}

\section{Mapping between BHPT and NR linear momentum}
\label{sec:mapping_linmom}
Next, we focus on the linear momentum. Similar to Section~\ref{sec:angmom_mapping}, we calculate the linear momentum directly from the waveforms $h^{\ell m}(t)$. We first compute the time derivative of the radiated linear momenta as~\cite{Gerosa:2018qay}:
\begin{align}
\frac{d P_x}{dt} = &\lim_{r \to \infty} \frac{r^2}{8\, \pi} \Re \Bigg[ \sum_{\ell,m} \,
\dot h^{\ell m}  \Big( a_{\ell m}\, \dot{\bar{h}}^{\ell,m+1}
 \notag\\
&+ b_{\ell,-m} \,\dot{\bar{h}}^{\ell-1,m+1}  -  b_{\ell+1,m+1}\, \dot{\bar{h}}^{\ell+1,m+1} \Big)\Bigg] \; ,
\label{eq:dt_px} \\
\frac{d P_y}{dt} = &\lim_{r \to \infty} \frac{r^2}{8\, \pi}\Im \Bigg[  \sum_{\ell,m}\, \dot h^{\ell m}  \Big( a_{\ell m}\, \dot{\bar{h}}^{\ell,m+1}
 \notag\\
&+ b_{\ell,-m} \,\dot{\bar{h}}^{\ell-1,m+1}  -  b_{\ell+1,m+1}\, \dot{\bar{h}}^{\ell+1,m+1} \Big)\Bigg] \; ,
\label{eq:dt_py}
\end{align}
\begin{align}
\frac{d P_z}{dt} = &\lim_{r \to \infty} \frac{r^2}{16 \pi} \sum_{\ell,m}\,
\dot{{h}}^{\ell m}   \Big( c_{\ell m}\, \dot{\bar{h}}^{\ell m}
\notag\\
&+ d_{\ell m}\, \dot{\bar{h}}^{\ell-1,m} +  d_{\ell+1,m}\, \dot{\bar{h}}^{\ell+1,m} \Big) \; ,
\label{eq:dt_pz}
\end{align}
where the upper bar denotes complex conjugation and the coefficients $a_{\ell,m}$, $b_{\ell,m}$, $c_{\ell,m}$, and $d_{\ell,m}$ are given in Ref.~\cite{Gerosa:2018qay}.
Integrating $d\mathbf{P}/dt$ then gives us the radiated linear momentum of the binary as a function of time:
\begin{equation}
    \mathbf{P}(t) = \int d\mathbf{P}/dt.
\end{equation}
While integrating, we set the integration constant to be zero as the linear momentum emission at early inspiral is expected to be negligible when averaged over an orbital cycle. Finally, we compute the total linear momentum as: $P=|\mathbf{P}|$.
We denote the computed NR and BHPT linear momentum as $P_{\tt NR}$ and $P_{\tt BHPT}$ respectively. The mass-scales are set to $M$.

\begin{figure}
\includegraphics[width=\columnwidth]{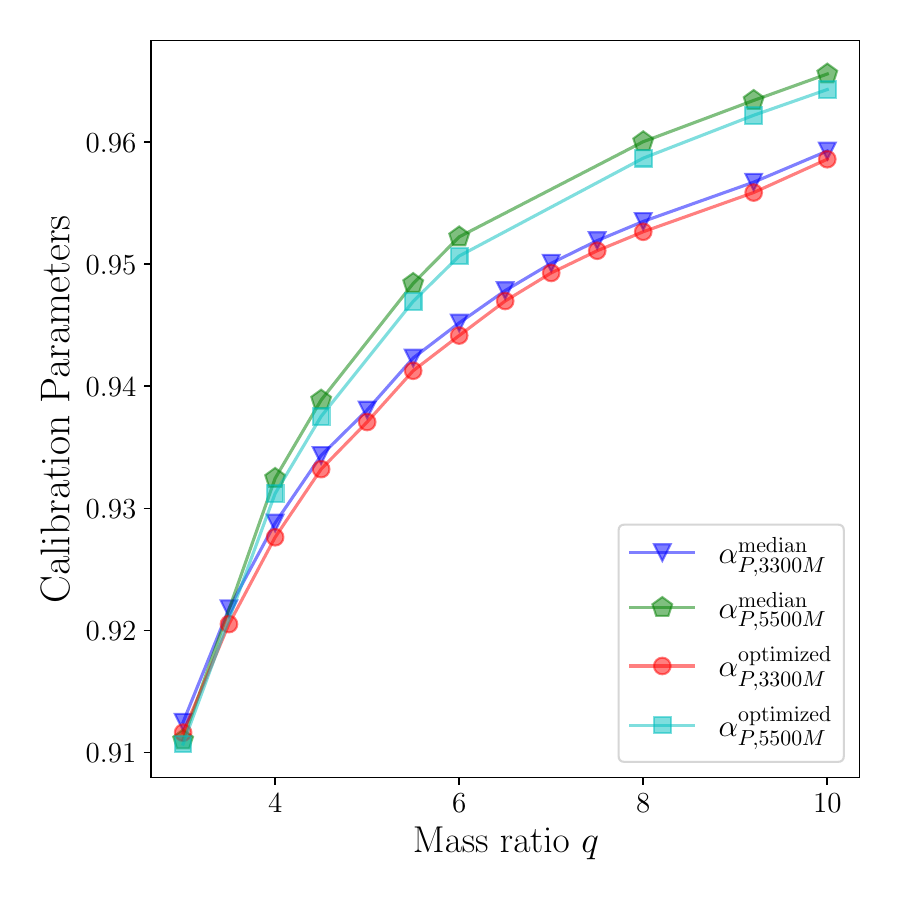}
\caption{We show the scaling parameters $\alpha_{P}$ (in particular, $\alpha^{\rm optimized}_{P,5500M}$, $\alpha^{\rm median}_{P,5500M}$, $\alpha^{\rm optimized}_{P,3300M}$ and $\alpha^{\rm median}_{P,3300M}$), obtained using different length of the NR data and methods as a function of the mass ratio ranging from $q=3$ to $q=10$. More details are in Section \ref{sec:mass_ratio_dependence_linmom}.}
\label{fig:alpha_linear_momentum_vs_q}
\end{figure}

\subsection{$\alpha$-$\beta$ scaling of the linear momentum}
\label{sec:alpha_beta_linmom}
We find that, for a given mass ratio, BHPT and NR linear momentum values are exhibit the following $\alpha$-$\beta$ mapping:
\begin{equation}
    P_{\tt NR} (t_{\tt NR}) =  \alpha_{P} \times P_{\tt BHPT} (\beta_{P} \times t_{\tt BHPT}),
    \label{eq:linmom_scaling}
\end{equation}
where $\alpha_{\mathcal{P}}$ and $\beta_{\mathcal{P}}$ are the scaling parameters. We adopt the same approach as described in Section~\ref{sec:alpha_beta_angmom} to compute $\alpha_{\mathcal{P}}$ and $\beta_{\mathcal{P}}$. Specifically, we set $\beta_{\mathcal{P}}=\beta_{J}=\beta_{\mathcal{E}}=\beta \times (1+1/q)$. Subsequently, we calculate an optimized value for $\alpha_{\mathcal{P}}$, denoted as $\alpha_{P}^{\rm optimized}$, and a median value of the expected $\alpha_P$ time series, denoted as $\alpha_{P}^{\rm median}$.

\subsection{Demonstration at $q=4$}
\label{sec:linmom_q4}
Following similar steps as in Section~\ref{sec:angmom_q4}, we demonstrate the linear momentum mapping at $q=4$ (Fig.\ref{fig:q4_linmom}). We observe that the differences between BHPT and NR linear momentum are much smaller compared to the angular momentum case. Additionally, the linear momentum values cover a vast range. Therefore, we present the linear momentum values on a logarithmic scale. Similar to what we observed in the angular momentum case, there are some differences between BHPT and NR linear momentum values. To emphasize the nature of these differences, we zoom into the binary evolution for $-4000M \leq t \leq -100M$ (Fig.\ref{fig:q4_linmom}, right panel inset). We note that the $\alpha_P$-$\beta_P$ scaled BHPT linear momentum matches the NR data very well in the inspiral part before the scaling breaks down in the merger-ringdown part.

\subsection{Mass ratio dependence}
\label{sec:mass_ratio_dependence_linmom}
Similar to the angular momentum case, following the procedure outlined in Section~\ref{sec:mass_ratio_dependence_angmom}, we repeat the experiment for another 12 mass ratio values ranging from $q=3$ to $q=10$ using two different length of common time grids, $\sim5500M$ and $\sim3300M$. This provides us with $\alpha_{P}^{\rm optimized}$ and $\alpha_{P}^{\rm median}$ for all these mass ratios. Additionally, we use $\sim5500M$ and $\sim3300M$ in the subscript to denote the length of NR data used. We show the mass ratio dependence of $\alpha_{P,5500M}^{\rm optimized}$, $\alpha_{P,5500M}^{\rm median}$, $\alpha_{P,3300M}^{\rm optimized}$ and $\alpha_{P,3300M}^{\rm median}$ in Figure~\ref{fig:alpha_linear_momentum_vs_q}. We observe that the value of these scaling parameters increase as mass ratio increases.

Finally, we obtain analytical expression of $\alpha_{P}$ (for which we use $\alpha^{\rm optimized}_{P,3300M}$) by performing fits in terms of $q$ (and $\nu$) (using the \texttt{scipy.optimize.curve\_fit}~\cite{scipyfit} module) and obtain:
\begin{align} 
\alpha^{\rm median}_{P}(q) \approx & 1 -0.08210918  \times (\frac{1}{q}) +0.02604475 \times (\frac{1}{q})^2\notag\\
& -0.00304274 \times (\frac{1}{q})^3 + 0.00012195 \times (\frac{1}{q})^4
\end{align}
and
\begin{align}
\alpha^{\rm median}_{P}(\nu) \approx & 1 -0.81606208 \times \nu + 5.84429252 \times \nu^2\notag\\
& -28.04024705 \times \nu^3 +35.61917356 \times \nu^4.
\end{align}

\begin{figure}
\includegraphics[width=\columnwidth]{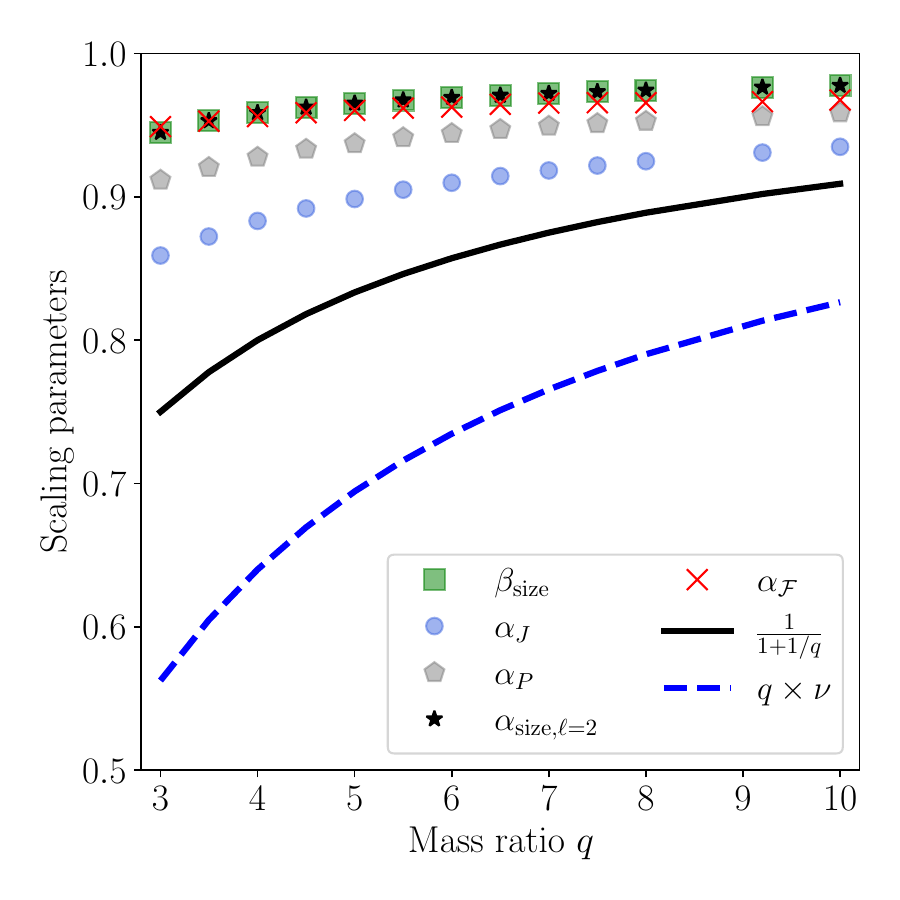}
\caption{We show the scaling parameters $\alpha$, obtained from $\alpha$-$\beta$ scalings for the waveform, fluxes, angular momentum, and linear momentum, and compare them to each other as a function of the mass ratio ranging from $q=3$ to $q=10$. More details are in Section \ref{sec:understanding_mapping}.}
\label{fig:alpha_vs_q}
\end{figure}

\section{Understanding the mapping}
\label{sec:understanding_mapping}
The first aspect we  explore is how the $\alpha$ values from different $\alpha$-$\beta$ scalings for the waveform, fluxes, angular momentum, and linear momentum compare to each other. In Figure~\ref{fig:alpha_vs_q}, we present both the $\alpha_{J}$ (obtained from the angular momentum scaling, blue circles) and $\alpha_{P}$ (obtained from the linear momentum scaling, grey pentagons) as functions of the mass ratio ranging from $q=3$ to $q=10$. For comparison, we also include the $\alpha_{\mathcal{F}}$, which scales the BHPT fluxes to match NR data (red circles)~\cite{Islam:2023csx}. Additionally, we show the $\beta_{\rm size}$ (green squares) and $\alpha_{\rm size,\ell=2}$ (black stars) from Ref.~\cite{Islam:2023qyt}, which are associated with the missing finite size of the smaller black hole in the BHPT framework. 

\begin{figure}
\includegraphics[width=\columnwidth]{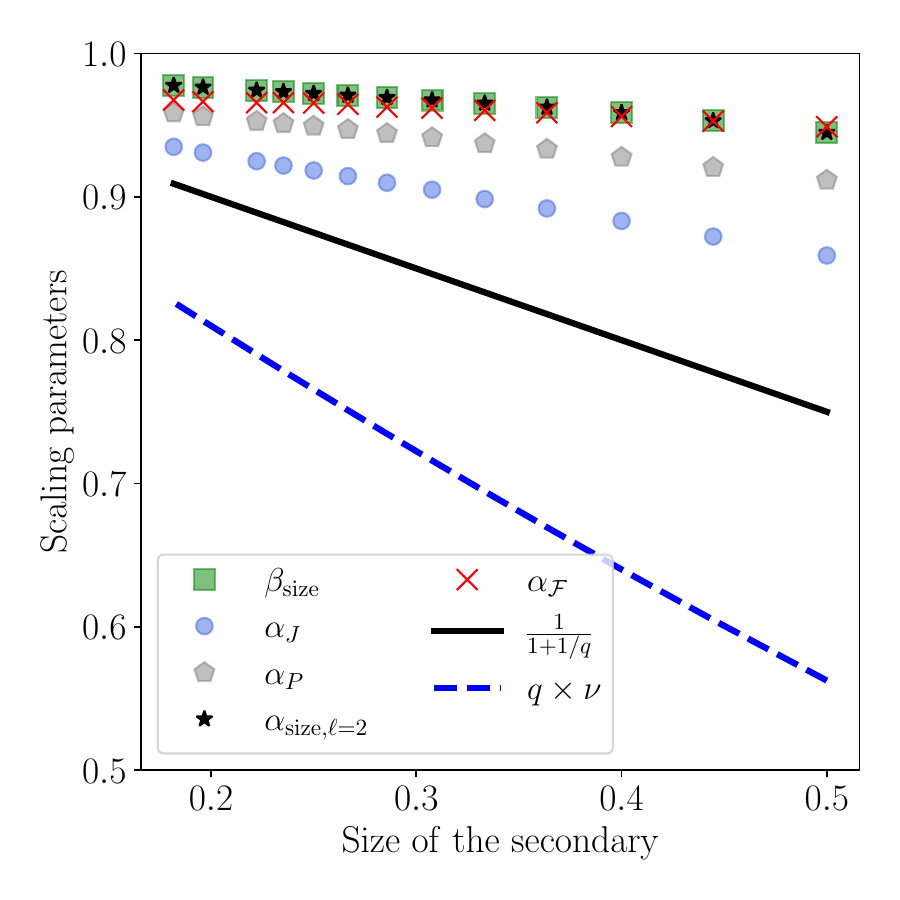}
\caption{We show the scaling parameters $\alpha$, obtained from $\alpha$-$\beta$ scalings for the waveform, fluxes, angular momentum, and linear momentum, and compare them to each other as a function of the radius of the smaller black hole for mass ratio ranging from $q=3$ to $q=10$. More details are in Section \ref{sec:understanding_mapping}.}
\label{fig:alpha_vs_rs}
\end{figure}

Finally, for context, we also include the mass-scale transformation factor $\frac{1}{1+1/q}$ from the mass-scale of $m_1$ to $M$, alongside the mass ratio transformation factor $q\nu$. Notably, these two transformation factors are considerably smaller than the $\alpha$ and $\beta$ values obtained from different scalings. Since we have already employed the same mass-scale for NR and BHPT, the mass-scale transformation factor is not required. Our BHPT framework, however, employs the mass ratio parameter $q$. If any of the $\alpha$ values were close to $q\nu$, it would have suggested that they primarily account for a more suitable mass ratio parameter ($\nu$).
It is interesting to observe that the $\alpha_P$, $\alpha_{\mathcal{F}}$, $\beta_{\rm size}$, and $\alpha_{\rm size,\ell=2}$ values closely align, suggesting a potential shared origin. On the other hand, $\alpha_P$ is notably smaller than the other $\alpha$ or $\beta$ values considered but slightly larger than the mass-scale transformation factor $\frac{1}{1+1/q}$.
We note here that while waveform or linear momentum have the unit of $M$, angular momentum has the mass-scale of $M^2$. This means that if we take the square root of $\alpha_J$, it should approach the other $\alpha$ or $\beta$ values. We confirm that this is indeed true.

\begin{figure}
\includegraphics[width=\columnwidth]{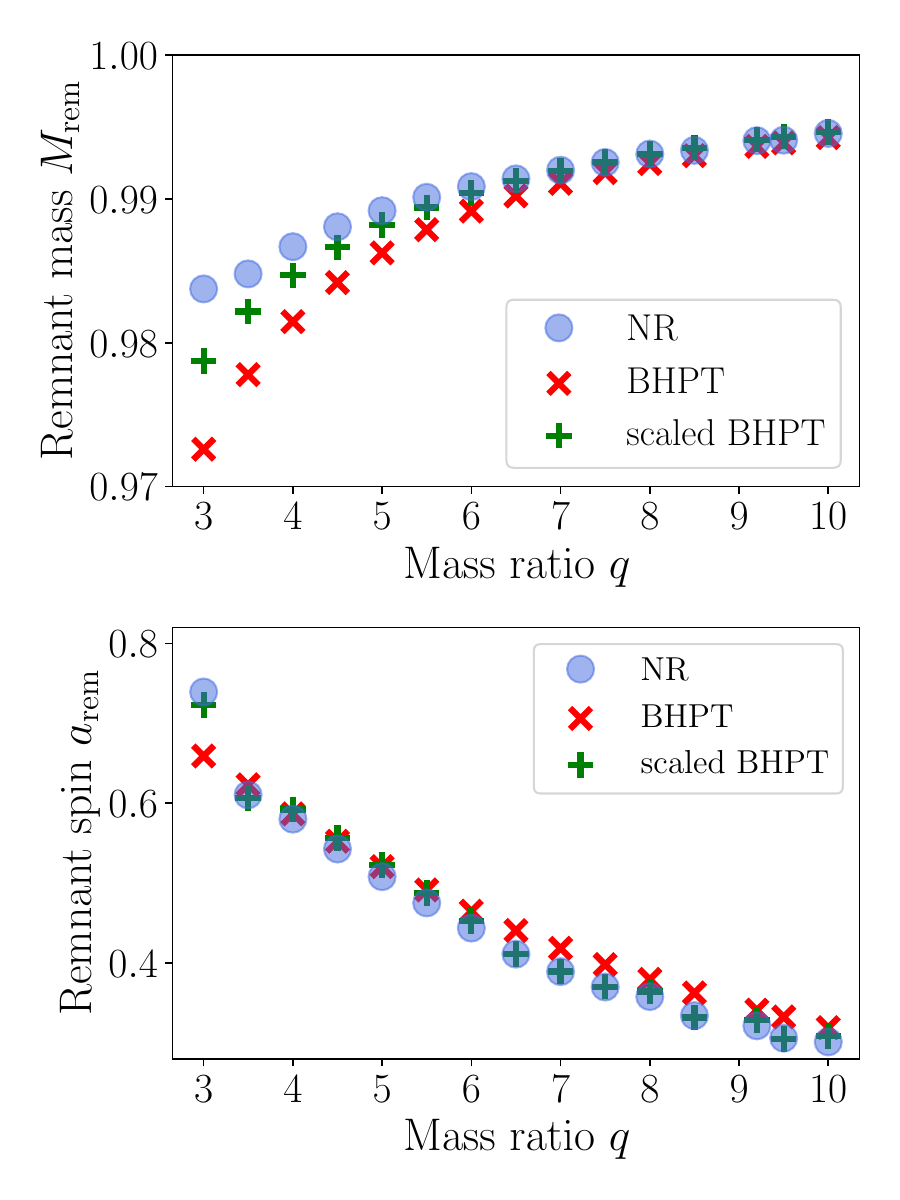}
\caption{We show the remnant mass $M_{\rm rem}$ and remnant spin $a_{\rm rem}$ as a function of the mass ratio calculated from both NR (blue circles) and BHPT (red crosses). Furthermore, we show the remnant properties computed using various $\alpha$-$\beta$ type scalings in green markers. More details are in Section \ref{sec:remnant}.}
\label{fig:remnant}
\end{figure}

\begin{figure}
\includegraphics[width=\columnwidth]{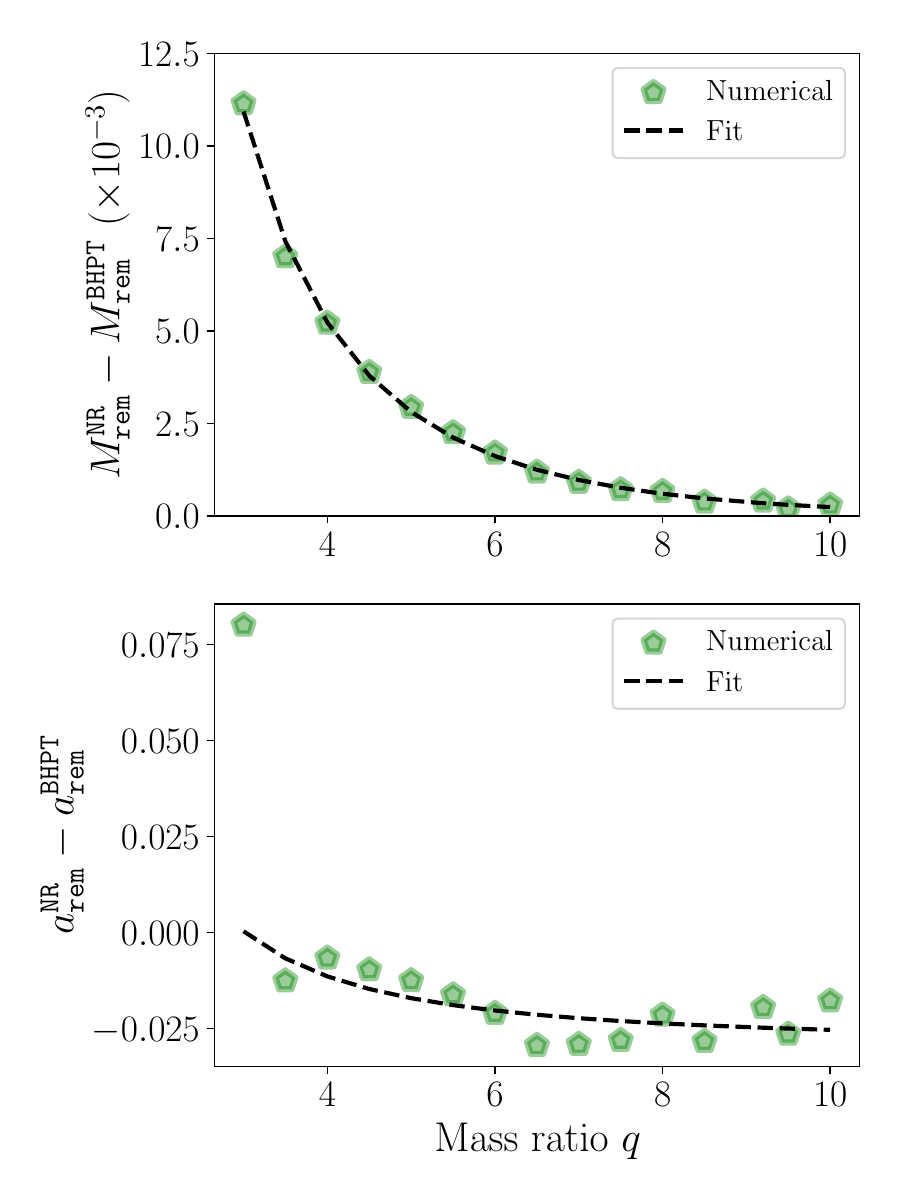}
\caption{We show the differences between in remnant mass (upper panel) and remnant spin (lower panel) computed using NR and BHPT waveforms as a function of the mass ratio. We also show the analytical fit used to capture these differences as black dashed lines. More details are in Section \ref{sec:remnant}.}
\label{fig:delta_remnant}
\end{figure}

To understand the relation between the missing finite size effect in BHPT and $\alpha$-$\beta$ scaling, we first compute the horizon area $\mathcal{A}$ of the secondary black hole as
\begin{equation}
\mathcal{A} = 8\pi \left( \frac{Gm_2}{c^2}\right)^2 (1 + \sqrt{1-\chi_2^2}),
\end{equation}
where $\chi_2$ is the spin of the secondary. For us, $\chi_2=0$. The horizon radius of the smaller black hole is then:
\begin{equation}
r_{\rm S} = \sqrt{\frac{\mathcal{A}}{4\pi}} = \frac{2}{1+q}.
\label{eq:radius}
\end{equation}
In Figure~\ref{fig:alpha_vs_rs}, we illustrate how different scaling parameters vary with the size of the smaller black hole. We observe that as the size becomes smaller, the corrections also become smaller, implying that they deviate less from unity.

\section{Implication on remnant quantities}
\label{sec:remnant}
Our analysis has interesting implications for the remnant properties. Below, we discuss how these energy and momentum scalings translate to the remnant quantities.

The remnant mass of a binary black hole merger is given by:
\begin{align}
M_{\rm rem} & = M_{\rm initial} - E_{\rm b} - \Delta M,
\label{eq:Mrem}
\end{align}
where $M_{\rm initial}$ and $E_{\rm b}$ are the total mass and binding energy of the binary at the start of the simulation. On the other hand, the total radiated mass until the merger is given by:
\begin{equation}
    \Delta M = \int \mathcal{F}(t') \, dt'.
\end{equation}
Using the flux scaling given in Eq.(\ref{eq:flux_scaling}), we can show that the radiated mass in BHPT and NR simulations is related by:
\begin{equation}\label{eq:mfcorr}
    \Delta M_{\rm NR} = \alpha_{\mathcal{F}} \beta \Delta M_{\tt BHPT}.
\end{equation}
Since $M_{\rm initial}$ and $E_{\rm b}$ will be the same for both NR and BHPT frameworks, the difference between NR and BHPT estimations of the remnant mass will be
\begin{equation}
    M_{\rm rem, NR} - M_{\rm rem, BHPT} \approx (\alpha_{\mathcal{F}} \beta - 1) \Delta M_{\tt BHPT}.
\end{equation}
Both $M_{\rm initial}$ and $E_{\rm b}$ will typically be much larger than $\Delta M_{\tt BHPT}$. Furthermore, $(\alpha_{\mathcal{F}} \beta - 1)$ will be approximately $-0.25$ for $q=4$ and would be closer to zero at higher mass ratios. This implies that the remnant mass estimated from NR and BHPT frameworks will not be extraordinarily different.

Similarly, the remnant spin is computed as:
\begin{eqnarray}
a_{\rm rem} = \frac{J^0_z-\Delta J_z}{M_{\rm rem}^2}\;,
\label{remnantspin}
\end{eqnarray}
where $J^0_z$ is the initial angular momentum and $\Delta J_z$ is the change of angular momentum until merger. Given, remnant mass between NR and BHPT are not noticeably different and $J^0_z$ is same for both NR and BHPT, any differences in the remnant spin estimation will show up due to the differences in $\Delta J_z$, which we compute as:
\begin{equation}
    \Delta J_z = \int J_z(t') \, dt'.
\end{equation}
Following Eq.(\ref{eq:angmom_scaling}), we can quickly write:
\begin{equation}
    \Delta J_{z,\rm NR} = \alpha_{J} \beta_{\rm size} \Delta J_{z,\tt BHPT}
\end{equation}
and
\begin{equation}\label{eq:sfcorr}
    \Delta J_{z,\rm NR} - \Delta J_{z,\rm BHPT} \approx (\alpha_{J} \beta_{\rm size} - 1) \Delta J_{z,\tt BHPT}.
\end{equation}
As, $\alpha_{J} \beta_{\rm size} \Delta J_{z,\tt BHPT} \ll J_z^0$ and $M_{\rm rem, NR} \approx M_{\rm rem, BHPT} $, differences between the remnant mass estimated using NR and BHPT will be close. Therefore, even though we will see differences between NR and BHPT momentum and fluxes and obtain effective scaling between them, these differences will not yield significant differences in the remnant mass or spin estimation. This provides some explanation of the observed match between NR and BHPT remnant spin values in Ref.~\cite{Price:2022lcx}. 

In Figure.~\ref{fig:remnant}, we show the remnant mass and remnant spin as a function of the mass ratio computed from both NR and BHPT. For the remnant spin, we find the values are pretty close to each other whereas for the remnant mass, initially we see some differences between NR and BHPT values. However, these differences are not large and more or less on the ball park. As the mass ratio increases, these differences decrease.
We further show that when various $\alpha$-$\beta$ scalings are incorporated through Eq.(\ref{eq:mfcorr}) and Eq.(\ref{eq:sfcorr}), computed remnant values match NR more closely. Specifically, computed remnant spin matches NR values very well for $q\ge4$ whereas remnant mass match NR values for $q \ge 5$. For $q \le 5$, we still see visible differences in remnant mass between NR and $\alpha_{\mathcal{F}}$-$\beta_{\mathcal{F}}$ scaled BHPT. However, these differences are smaller than the difference between NR and BHPT. 

Differences seen for $q \le 5$ possibly indicate modelling inaccuracies of the $\alpha$-$\beta$ scalings for these mass ratios. It will be interesting to explore higher order corrections in the $\alpha$-$\beta$ scalings near equal mass limit. As a first step, we do the following. We compute the differences in computed remnant mass and spin values obtained from NR and BHPT waveforms, and fit them with a polynomial in $\frac{1}{q}$ (using the \texttt{scipy.optimize.curve\_fit}~\cite{scipyfit} module). We obtain:
\begin{align} 
M_{\rm rem}^{\rm NR} - M_{\rm rem}^{\rm BHPT} \approx 0.00068 + \frac{-0.01953}{q} + \frac{0.15083}{q^2},
\end{align}
and
\begin{align}
a_{\rm rem}^{\rm NR} - a_{\rm rem}^{\rm BHPT} \approx -0.02947 + \frac{0.02043}{q} + \frac{0.20659}{q^2}.
\end{align}
We find that a second order polynomial in $\frac{1}{q}$ nicely captures the difference in remnant mass estimated using NR and BHPT waveforms (Figure.~\ref{fig:delta_remnant}). For the remnant spin, the difference do not show a nice trend. However, a second-order polynomial in $\frac{1}{q}$ can capture the overall trend to some extent. 

\section{Concluding remarks}
\label{sec:conclusion}
In this paper, we unveil two new scalings between the linear BHPT solution and fully nonlinear NR data in the comparable mass regime for non-spinning binaries. Our findings contribute additional evidence that these seemingly distinct frameworks are more closely interconnected than previously thought. Besides the previously reported $\alpha$-$\beta$ scaling for waveform modes and fluxes, we identify similar scalings for linear and angular momentum. What makes this finding intriguing is that the linear and angular momentum of gravitational waves are intricate products of various waveform modes and their time derivatives, making it challenging to anticipate whether the existence of a scaling between waveform modes would extend to radiated overall quantities like angular and linear momentum. 

For completeness, we list all the $\alpha$-$\beta$ scalings below:
\begin{align}
h^{\ell,m}_{\tt NR}(t_{\tt NR} ; q) \sim {\alpha_{\ell,\rm size}} h^{\ell,m}_{\tt BHPT}\left( \beta_{\rm size} t_{\tt BHPT};q \right) \,,\\
\mathcal{F}_{\tt NR} (t_{\tt NR}) =  \alpha_{\mathcal{F}} \mathcal{F}_{\tt BHPT} (\beta_{\rm size} t_{\tt BHPT}),\\
{J}_{z,\tt NR} (t_{\tt NR}) =  \alpha_{{J}} {J}_{z,\tt BHPT} (\beta_{\rm size} t_{\tt BHPT}),\\
P_{\tt NR} (t_{\tt NR}) =  \alpha_{P} P_{\tt BHPT} (\beta_{\rm size} t_{\tt BHPT}).
\end{align}
Note that we use the same $\beta_{\rm size}$ for all these scalings. The value of $\beta_{\rm size}$ is related to $\beta$ (by a mass-scale transformation factor of $\frac{1}{1+1/q}$) used for the waveform. 

Furthermore, we have provided some explanation why these scalings, or in general, the differences between NR and BHPT does not affect the remnant property estimation significantly. However, we do find that when these scalings are incorporated, BHPT and NR remnant properties match more closely than before.

We believe that these scalings will illuminate the interplay between the NR and BHPT frameworks in the coming days, providing insights into a deeper understanding of binary black hole dynamics. While there are indications that these scalings are linked to the missing finite size of the smaller black hole~\cite{Islam:2023aec}, more systematic studies are necessary to fully address this issue. Additionally, it is crucial to investigate whether these scalings extend to more complex scenarios involving spins and eccentricities. We aim to explore and answer these questions in the near future.

\begin{figure}
\includegraphics[width=\columnwidth]{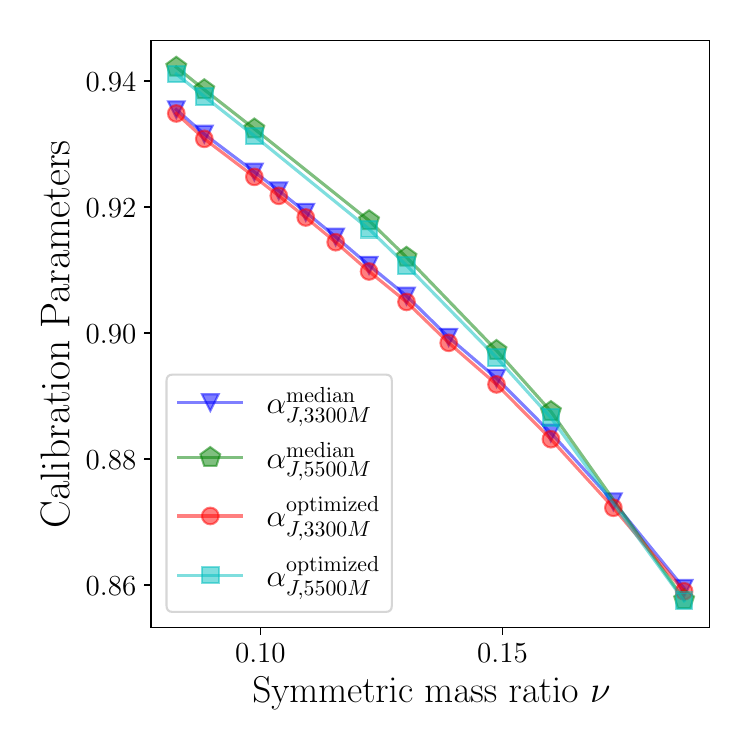}
\caption{We show the calibration parameters $\alpha_{J}$ (in particular, $\alpha^{\rm optimized}_{J,5500M}$, $\alpha^{\rm median}_{J,5500M}$, $\alpha^{\rm optimized}_{J,3300M}$ and $\alpha^{\rm median}_{J,3300M}$), obtained using different length of the NR data and methods as a function of the symmetric mass ratio $\nu$. More details are in Section \ref{sec:nu}.}
\label{fig:alpha_angular_momentum_vs_nu}
\end{figure}

\begin{figure}
\includegraphics[width=\columnwidth]{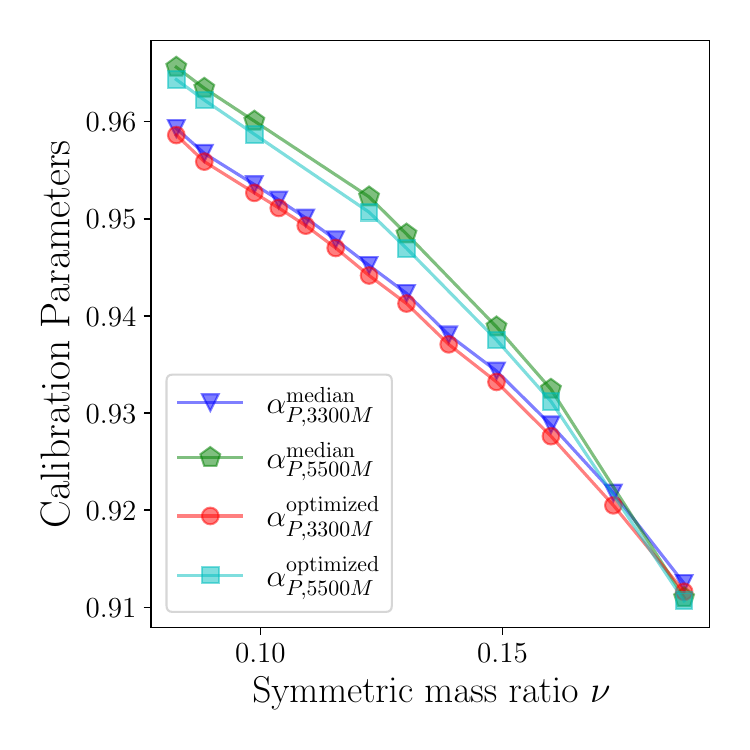}
\caption{We show the scaling parameters $\alpha_{P}$ (in particular, $\alpha^{\rm optimized}_{P,5500M}$, $\alpha^{\rm median}_{P,5500M}$, $\alpha^{\rm optimized}_{P,3300M}$ and $\alpha^{\rm median}_{P,3300M}$), obtained using different length of the NR data and methods as a function of the symmetric mass ratio $\nu$. More details are in Section \ref{sec:nu}.}
\label{fig:alpha_linear_momentum_vs_nu}
\end{figure}

\begin{acknowledgments}
We thank Gaurav Khanna and Scott Field for helpful discussions and thoughtful comments on the manuscript. We also thank the SXS collaboration for maintaining publicly available catalog of NR simulations which has been used in this study.
\end{acknowledgments}

\appendix 
\section{Mass ratio dependence of the scaling parameters}
\label{sec:nu}
In Figure~\ref{fig:alpha_angular_momentum_vs_nu} and ~\ref{fig:alpha_linear_momentum_vs_nu}, we show how the scaling parameters $\alpha_{J}$ and $\alpha_P$ changes as a function of the symmetric mass ratio $\nu$ corresponding to $3 \leq q \leq 10$. We show the scaling parameters computed using both $3500M$ and $5500M$ of NR data. Details of these parameters are given in Section~\ref{sec:mass_ratio_dependence_angmom} and Section.~\ref{sec:mass_ratio_dependence_linmom}.

\bibliography{References}

\end{document}